\documentclass[conference,compsoc,nocompress,letterpaper]{IEEEtran}

\usepackage{etoolbox}
\makeatletter
\patchcmd{\@makecaption}
  {\scshape}
  {}
  {}
  {}
\makeatother

\usepackage{usenix-2020-09}
\usepackage{cite}
\usepackage{amsmath,amssymb,amsfonts}
\usepackage{algorithmic}
\usepackage{graphicx}
\usepackage{textcomp}
\usepackage{xcolor}
\def\BibTeX{{\rm B\kern-.05em{\sc i\kern-.025em b}\kern-.08em
    T\kern-.1667em\lower.7ex\hbox{E}\kern-.125emX}}
\usepackage{multirow} 
\usepackage{threeparttable}

\begin{document}

\def\anon{0}
\def\ieeestylepaper{1}

\title{\Large \bf Pandora's Box in Your SSD: The Untold Dangers of NVMe}

\if\ieeestylepaper1
\if\anon1
\author{
\IEEEauthorblockN{Anonymous author list...}
\IEEEauthorblockA{Anonymous institution...\\
\vspace{0.7cm}}
}
\else
\author{
\IEEEauthorblockN{Rick Wertenbroek$^{1,2}$, and Alberto Dassatti$^1$}
\IEEEauthorblockA{$^1$\textit{School of Engineering and Management Vaud, University of Applied Sciences and Arts Western Switzerland}\\
$^2$\textit{University of Lausanne,}\\
rick.wertenbroek@heig-vd.ch, alberto.dassatti@heig-vd.ch \\}
}
\fi
\else
\author{
{\rm Anonymous Author}\\
Firsts Institution
\and
{\rm Anonymous Authors...}\\
Other Institutions...
} 
\fi

\maketitle

\begin{abstract}
Modern operating systems manage and abstract hardware resources, to ensure efficient execution of user workloads. The operating system must securely interface with often untrusted user code while relying on hardware that is assumed to be trustworthy. In this paper, we challenge this trust by introducing the eNVMe platform, a malicious NVMe storage device. The eNVMe platform features a novel, Linux-based, open-source NVMe firmware. It embeds hacking tools and it is compatible with a variety of PCI-enabled hardware. Using this platform, we uncover several attack vectors in Linux and Windows, highlighting the risks posed by malicious NVMe devices. We discuss available mitigation techniques and ponder about open-source firmware and open-hardware as a viable way forward for storage. While prior research has examined compromised existing hardware, our eNVMe platform provides a novel and unique tool for security researchers, enabling deeper exploration of vulnerabilities in operating system storage subsystems.

\end{abstract}

\if\ieeestylepaper1
\fi

\section{Introduction}

Non-Volatile Memory express (NVMe) is becoming the de-facto standard for fast Solid-State Drives (SSDs). Most computers sold nowadays come with one or more NVMe SSDs, which are thus ubiquitous. The NVMe standard is based on PCI Express (Peripheral Component Interconnect Express, or PCIe in short) a high-speed serial computer expansion bus. 
NVMe SSDs come equipped with multi-core embedded controllers and a number of Direct Memory Access (DMA) engines to handle the large number of Input-Output (IO) requests per second (IOPS) and are capable of bandwidths up to tens of gigabytes per second. Today's storage controllers are no longer \textit{dumb devices} and pack substantial compute power, often with specialized co-processors, crypto-engines, inline compression, and more. The capabilities of NVMe storage are growing by the day through efforts such as computational storage, backed by the NVMe standard itself~\cite{nvme2024cs}, and the Storage Networking Industry Association (SNIA) technical works on computational storage~\cite{snia2024cs}.
The combination of this processing power and the strategic position in the hardware architecture of a computer, make NVMe SSDs the perfect vector for large-scale attacks and a cyber-warfare super-weapon. To the best of our knowledge, storage-based attacks remain a largely unexplored threat vector.

In this work, we try to raise awareness of the serious threats that an \textit{evil} NVMe device can pose to the global computing infrastructure. To this end, we designed a low-cost, Linux based, NVMe capable research platform. We implanted this platform in several hardware machines running Linux and Windows Operating Systems (OS) and used it to take full control of our targets. We started with simple DMA attacks when the victim's hardware configuration allowed it. When this was not possible, we used the device's storage capabilities to develop new attack vectors. 

We open this work with a cautionary tale depicting possible scenarios of weaponized NVMe SSDs. We hope it motivates the necessity of a research platform to study these scenarios; After the story, we provide the necessary background and introduce our research platform. We then demonstrate a number of attacks that make the initial story-line less fictional than it may appear at first glance. Thereafter, we discuss mitigation techniques and their limits. Finally, we conclude with a call to action for a transparent storage system initiative. Our main contributions in this paper are the following:
\begin{itemize}
    \item A low-cost and fully open-source platform to explore the security implications of an \textit{evil} NVMe device;
    \item A demonstration that many systems today are vulnerable to storage-specific attacks;
    \item A number of reproducible attacks;
    \item Storage-specific attacks demonstrating that, even if a target computer is properly configured and protected against traditional DMA attacks, it is still vulnerable from its internal storage.
\end{itemize}

\noindent
Available on GitHub: \url{https://github.com/rick-heig/eNVMe}

\subsection{A cautionary tale}
\label{sec:caution}

Let's imagine government \textbf{X} has control over company \textbf{$\Phi$}. \textbf{$\Phi$} is a well-respected NVMe controller chip maker. \textbf{$\Phi$}'s controllers are used by big storage brands all around the world. These storage brands have been selling NVMe drives with \textbf{$\Phi$}'s controllers for several years and have a large market share: they can be found in most laptops, workstations, and data centers. It turns out that government \textbf{X} tasked \textbf{$\Phi$} to add some extra capabilities and dormant code inside its controllers.

A war breaks out and government \textbf{X} made enemies all around the world. In face of this global conflict, government \textbf{X} decides to unleash its weapon, an army of NVMe SSDs!

\textbf{X} does not have direct remote access to the NVMe SSDs, however, let's say it decides to launch a promotional ad campaign on the Internet for a trendy online marketplace. The ad campaign uses cookie technology for tracking. The tracking cookie contains a unique binary key. People from all around the world see the ads while browsing popular sites or when using their search engine. The cookie gets stored on the NVMe SSD. It turns out the unique key value in the cookie is actually the \textit{nuclear button} for the NVMe SSDs to turn \textit{evil}, apocalypse ensues...

Although this is an imaginary tale, there have been multiple reports of firmware poisoning~\cite{cui2013firmware, hudson2015thunderstrike, bettayeb2019firmware}, intentionally malicious hardware~\cite{bhunia2014hardware, shwartz2017shattered, blanchet2018badusb, salmani2018global}, and governmental implication in the development of hardware cyber-warfare and cyber-security tools~\cite{farwell2011stuxnet, fitzpatrick2014stupid, gallagher2017new, durden2017wikileaks}. Therefore, this might actually be unfolding right now, behind our backs, and dormant malicious NVMe SSDs may already be installed in a number of PCs.

\subsubsection{The apocalypse}

\subsubsection*{Death}

Upon being triggered by the activation unique key, all NVMe SSDs with a \textbf{$\Phi$} controller destroy themselves. They will never start again, all their data is erased, the chip destroyed itself, millions of computers are non-operational. This is, however, quite brutal and easily detectable. While a possible attack, other more subtle and powerful approaches may be devised.

\subsubsection*{Conquest}

Upon activation, the SSDs take control of their host operating systems, they now have total control of millions of machines, all conquered, silently.

\subsubsection*{Famine}

The \textbf{$\Phi$}-equipped SSDs have been scanning their own contents in the background for years, they know people and companies have been storing their private access keys, cryptocurrency wallets, seed phrases, etc. on the SSD. After conquest, all wallets and keys are sent back to a given IP address controlled by \textbf{X}. Government \textbf{X} is now the de-facto cryptocurrency king! All wallets and funds are moved in one go, the victims are now starving and \textbf{X} possesses a new gigantic war fund.

\subsubsection*{War}

The SSDs can spy on all files stored on it, but also in most cases all files on the computer, not only the files inside the SSD. The SSD has been mining for interesting, secret, documents, intelligence of any kind, it can silently transfer them to one of the attacker servers. Besides spying, the SSDs can also alter any content, inject malware, and use the host machine for its own \textit{evil} schemes.

\subsection{Motivation}

Although the imagined scenarios sound right out of a science-fiction thriller, they can be very close to reality. Our goal with this work is to raise awareness about how vulnerable computers are from storage devices, and provide a platform to study NVMe-based attacks. Most of the state-of-the-art research in NVMe SSD security is focused on the software and driver side and very little research is available on the device itself being the attack vector. The National Institute of Standards and Technology (NIST) released security guidelines for storage infrastructure~\cite{chandramouli2020security} that formalizes the threats, risks, and attack surface while discussing security guidelines for the storage infrastructure. The document lists the following threats: Credential theft or compromise, Cracking encryption, Infection of malware and ransomware, Backdoors and unpatched vulnerabilities, Privilege escalation, Human error and deliberate misconfiguration, Physical theft of storage media, Insecure images, software and firmware.

To the best of our knowledge, there is no security-oriented publicly available NVMe hardware research platform for testing these kinds of attacks from the disk itself rather than from within the storage infrastructure. In this work we fill this gap by providing a fully open-source platform for the development of NVMe compliant hardware focused on security and penetration testing.

\section{Background}

Before we introduce the NVMe research platform and demonstrate how to implement the attacks of the cautionary tale we provide the background on how NVMe functions and why NVMe SSDs could pose such as threat.

\subsection{NVMe}

Non-Volatile Memory Express (NVMe) is an open and widely adopted interface specification for accessing non-volatile storage~\cite{nvme2022base, nvme2022transportpcie}. NVMe was originally designed to standardize high-end SSD interfaces using PCIe. NVMe eliminates the need for vendor-specific drivers, akin to USB mass storage devices that follow standard specifications. The NVMe standard covers a variety of form factors and interfaces, from PCIe~\cite{pci2021pci} to M.2~\cite{pci2024m2} and U.2~\cite{pci2024u2}. NVMe/PCIe can also be tunneled over Thunderbolt~\cite{merrit2011pci} or USB-C~\cite{ziller2015thunderbolt, byrne2015one}. More recently, even some SD cards, rely on NVMe. They are equipped with an on-card PCIe interface and comply with the SD Express standard~\cite{sd2020express}. Thunderbolt, USB-C, and SD express slots make the ubiquitous NVMe even more accessible with direct external connectors.
NVMe enables high levels of parallelism through a queue submission mechanism. The queues allow the multiple cores of modern CPUs to interact with the device simultaneously. An overview of NVMe is illustrated in Figure~\ref{fig:nvme1}. The CPU prepares IO submissions (requests) in IO submission queues in RAM. These submissions describe the source and destination of the data. For write submissions, the source is in RAM and destination is an NVMe logical block address inside the disk. Similarly, for read submissions, the source is an NVMe logical block address inside the disk and the destination is in RAM. Once the CPU prepared the submissions in RAM, it writes a PCI register of the NVMe drive to indicate that there is work to do. The NVMe SSD then accesses RAM directly for both reading the submissions queue, reading data to store on disk, writing data retrieved from disk, and filling the completion queue. Once the submissions are completed the NVMe SSD sends an IRQ to the CPU so that it can check the completion queue. The main difference between NVMe and older SATA drives is that the DMA engine resides inside the disk. The DMA engines are entirely controlled by the NVMe SSD controller, which uses them to complete the IO submissions.

\subsection{Direct Memory Access (DMA)}

\begin{figure}
\begin{center}
\includegraphics[width=0.48\textwidth]{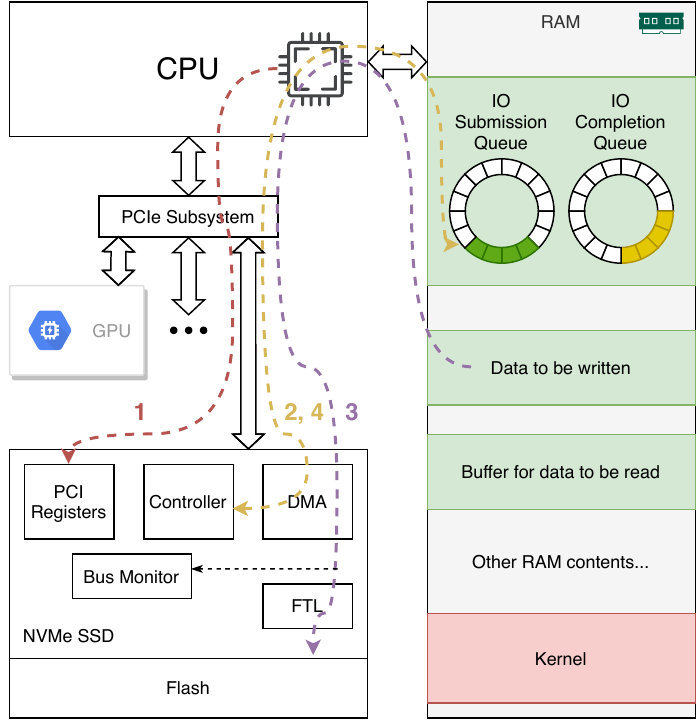}
\end{center}
\caption{\label{fig:nvme1} \textbf{Overview of NVMe write data example:} An NVMe SSD is connected over PCIe and has direct memory access to the system RAM. In order to write (or read) data to an NVMe SSD the CPU prepares IO submissions (requests) in an IO submission queue (SQ) in RAM. These submissions describe the source location of the data and destination inside the SSD. (\textbf{1}) The CPU rings the NVMe SSD's doorbell by writing a PCI register, letting the NVMe SSD know there is work to do. (\textbf{2}) The SSD controller fetches the submissions from the IO SQ in RAM. (\textbf{3}) The SSD reads data from RAM with DMA and stores it in flash. (\textbf{4}) The controller writes the completions in the IO completion queue (CQ) in RAM and sends an IRQ to the CPU to indicate the submissions have been completed. Read transactions are done in a similar manner but with data being moved from flash to a buffer in RAM. Green regions in RAM are supposed to be accessible to the NVMe SSD, gray and red regions are not supposed to be accessed by the NVMe SSD.}
\end{figure}

Direct memory access to a computer physical memory space is one of the highest levels of privilege and a prime vector for attacks. As DMA-enabled devices have complete access to the state of a computer, they can fully compromise it. DMA not only allows for access to the RAM but also allows to read and write other components inside the physical address space, as well as other devices on the PCIe bus. Although there are mitigations, such as memory protection by the IO Memory Management Unit (IOMMU), that would allow the NVMe SSD to only access the green parts in Figure~\ref{fig:nvme1}, they are not always enabled or properly configured~\cite{markettos2019thunderclap}. Without IOMMU the NVMe device can access everything in RAM, including the Kernel and other RAM contents as shown in Figure~\ref{fig:nvme1}.

An Input-Output (IO) Memory Management Unit (IOMMU) is a Memory Management Unit (MMU) connecting a DMA-capable IO device (here via PCIe) to the physical memory address space. Like a traditional MMU, which translates CPU virtual addresses to physical addresses, the IOMMU maps device (IO) virtual addresses (IOVA) to physical addresses.
Both main x86 CPU manufacturers, AMD and Intel have published specifications for their IOMMU technology. AMD's IOMMU technology is called AMD-Vi~\cite{amd2023technology}, and Intel's technology is called Virtualization Technology for Directed IO, VT-d~\cite{intel2022technology}. ARM CPUs also provide an IOMMU technology called System Memory Management Unit (SMMU)~\cite{arm2024technology}.
When enabled and properly configured, the IOMMU restricts the access of devices to certain physical pages. With IOMMU active, the NVMe device would only be able to access the RAM regions that it is supposed to, i.e., green regions in Figure~\ref{fig:nvme1}, and accesses to other RAM contents would result in an IO page fault. It is to note that devices in the same IOMMU group are not isolated from each other, for example, devices under the same PCIe bridge~\cite{redhat2024deep}.
While in theory IOMMU would restrict access and isolate the device avoiding access to unwanted locations,
we show ways to circumvent the IOMMU in section~\ref{sec:defiommu}. The page size granularity of the IOMMU can also lead to exposing some critical parts of memory~\cite{markettos2019thunderclap}. More alarmingly, section~\ref{sec:iommu} brings evidence that the IOMMU is, unfortunately, not properly configured or enabled on most systems.

\subsection{DMA hardware security considerations}
Direct memory access attacks rely on devices that have DMA capabilities. Several devices implement DMA to improve performance: examples include network adapters, GPUs, USB controllers, FireWire controllers, storage controllers (e.g., NVMe), accelerators, and programmable devices (e.g., FPGAs). Traditionally these devices relied on busses internal to the host computer, but more recently, FireWire, Thunderbolt and USB-C have made DMA possible also from outside the case of a computer. This high level of access gave birth to a large number of attacks~\cite{clark1974input, becher2005firewire, boileau2006hit, sang2011attacks, stewin2013understanding, sevinsky2013funderbolt, kalenderidis2014thunderbolts, frisk2016direct, gallagher2017new, neuner2017bad, trikalinou2017taking, blanchet2018badusb, delaunay2018practical}. The effectiveness of DMA-based attacks has led to the development of generic DMA attack platforms, such as SLOTSCREAMER~\cite{fitzpatrick2014stupid,slotscreamer2014}, INCEPTION~\cite{inception}, PCILeech~\cite{frisk2016direct, pcileech2024}, and Thunderclap~\cite{markettos2019thunderclap}. Beyond the attack platforms themselves, multiple frameworks were developed to facilitate the attacks~\cite{breuk2012integrating, leechcore2024}. IOMMU is widely credited to be the final solution against DMA attacks. While IOMMU is essential to prevent these attacks, alone it is unable to stop many attacks as previously demonstrated by \cite{markettos2019thunderclap}. Other researchers \cite{kim2023devious} have shown that even when IOMMUs are configured correctly and there are no software bugs, a DMA capable device can collect valuable information via side-channel attacks involving the IO Translation Look-aside Buffer (IOTLB). The interest around IOMMU attack space is fairly recent and the attack space not yet fully explored. As opposed to other hardware, NVMe SSDs combine three major features:First, they are DMA capable devices; Second, they hold data, and, finally, they have consequent compute capabilities. The synergies between these features lead to the powerful attacks that will be demonstrated in Section~\ref{sec:appattack}.

\section{The \textit{eNVMe} platform}
\begin{figure}
\begin{center}
\includegraphics[width=0.48\textwidth]{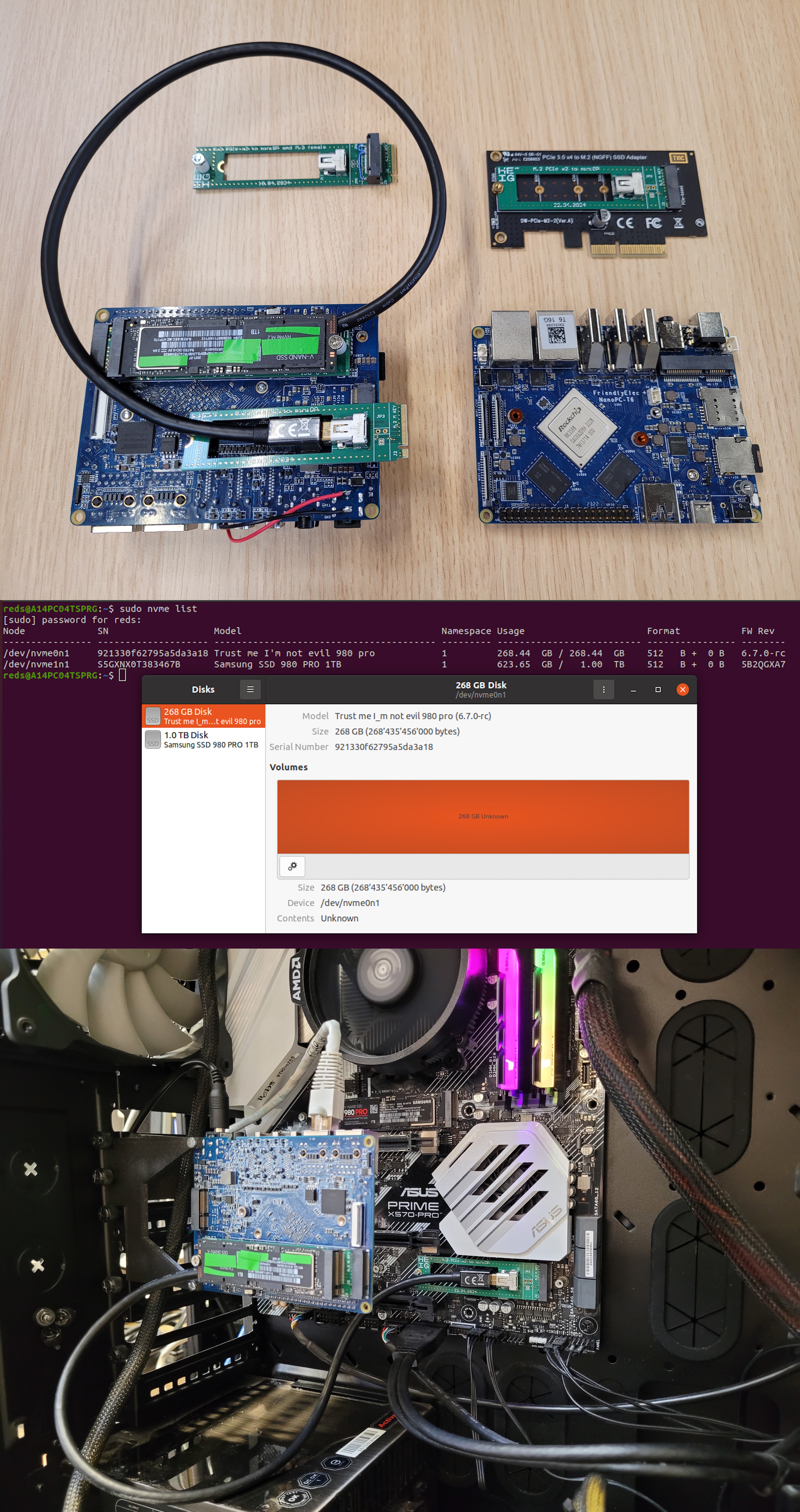}
\end{center}
\caption{\label{fig:t6} \textbf{Top:} Rockchip RK3588 SoC based single board computer (SBC) FriendlyElec NanoPC-T6 (bottom and top view). The T6 runs our NVMe firmware and comes with adapter PCBs to connect to target computer M.2 or regular PCI express interfaces. Unpopulated adapter PCBs are shown on top. \textbf{Middle:} a screenshot of our eNVMe platform inside the target host computer, where it is recognized as a regular NVMe SSD. We chose a model name that allows us to differentiate it, but we can choose any model name. \textbf{Bottom:} Our aNVMe platform mounted in a target host PC (bottom M.2 slot) alongside a regular NVMe SSD (top M.2 slot). Access to the T6 is done through an Ethernet cable and UART, but a screen keyboard and mouse can also be connected as it is a full-blown single board computer.}
\end{figure}

Before describing how eNVMe can be used to attack a system, let us describe what eNVMe is.  
The \textit{evil} NVMe platform\footnote{\if\anon1{\url{https://github.com/anonymous/project} (hidden for anonymous review)}\else{\url{https://github.com/rick-heig/eNVMe}}\fi}, or eNVMe in short, is an open-source NVMe firmware, operating system, and tools running on an embedded board. An incarnation of the platform is shown in Figure~\ref{fig:t6}. eNVMe enables researchers to study not only NVMe, but also allows the exploration of under-studied attack vectors linked to NVMe storage devices. The main goals of this platform are:
\begin{enumerate}
    \item Provide a physical hardware platform for NVMe security research.
      This is mandatory to study NVMe security scenarios in real-life settings and not rely on emulation or simulation.
    \item Compatibility: Implement the full NVMe standard and being recognized as a functional NVMe SSD by any host.
    \item Portability: Make the code run on a variety of hardware. We especially support affordable hardware that it is in reach for most researchers and students.
    \item Flexibility: 
    \begin{itemize}
        \item Allow the NVMe firmware to be modified freely.
        \item Support running C code, scripts, programs, or most advanced tools.
        \item Provide means to run user-defined code in the NVMe pathway or alongside it.
        \item Provide a sophisticated controller that allows interacting with its stored contents with minimal constraints.
    \end{itemize}
    \item Accessibility: we aim at making a customizable tool reducing the effort to prototype new attacks. 
\end{enumerate}

These objectives led to the platform and architecture described below. Our NVMe firmware is open-source. Furthermore, we chose to implement it as a Linux PCI endpoint which is a kernel module (driver) that relies on the Linux PCI endpoint framework\footnote{\url{https://docs.kernel.org/PCI/endpoint}} to implement a PCI function, in our case the NVMe function. Thanks to this framework, the driver is portable to different devices as long as they provide a driver for their PCI controller in endpoint mode (acting as a device as opposed to root-complex mode). Implementing NVMe as a PCI function in Linux ensures portability because any hardware that has a PCI endpoint controller, and associated driver, can now run our code. Relying on the Linux kernel also bring its rich user-space. Although our NVMe function runs in kernel, we implemented several communication channels to user-space in order to allow user-space code to interact with the NVMe code. These channels open a large spectrum of applications and allow using scripting languages, libraries, and any user-space software. We focused on making possible running our NVMe code on off-the-shelf, readily available and inexpensive Single Board Computers (SBCs). Whilst our code can run on sub 100\$ SBCs, this does not limit the possibilities to run on more advanced hardware and we do support Field Programmable Gate Array (FPGA) SoC platforms such as the AMD ZynqUltraScale+ family. This opens up the possibility to add custom hardware, implemented in the FPGA, between the PCI controller and the NVMe logic. Our firmware also allows the development NVMe computational storage devices~\cite{wertenbroek2024portable}.

With the eNVMe platform we can demonstrate the reality behind the cautionary tale of section~\ref{sec:caution} and attack real computers in realistic settings. It is true that this approach could also be done in emulation, e.g., through QEMU~\cite{bellard2005qemu, bartholomew2006qemu} for instance, but it would be limited to the emulated hardware and would not allow, for example, to take advantage of bugs in actual hardware. On real hardware equipped with security devices, such as input-output memory management units (IOMMU) and Trusted Platform Modules (TPM) we can explore the actual interactions that would occur between devices. As our NVMe firmware is Linux-based, it allows to implement many high-level features inside our NVMe drive with ease. One example is file-system awareness of the contents stored on the NVMe SSD; Another one is the ease of running user space programs, for instance the large number of exploits available in Kali Linux~\cite{hertzog2017kali, cisar2019some}.

\begin{figure}
\begin{center}
\includegraphics[width=0.48\textwidth]{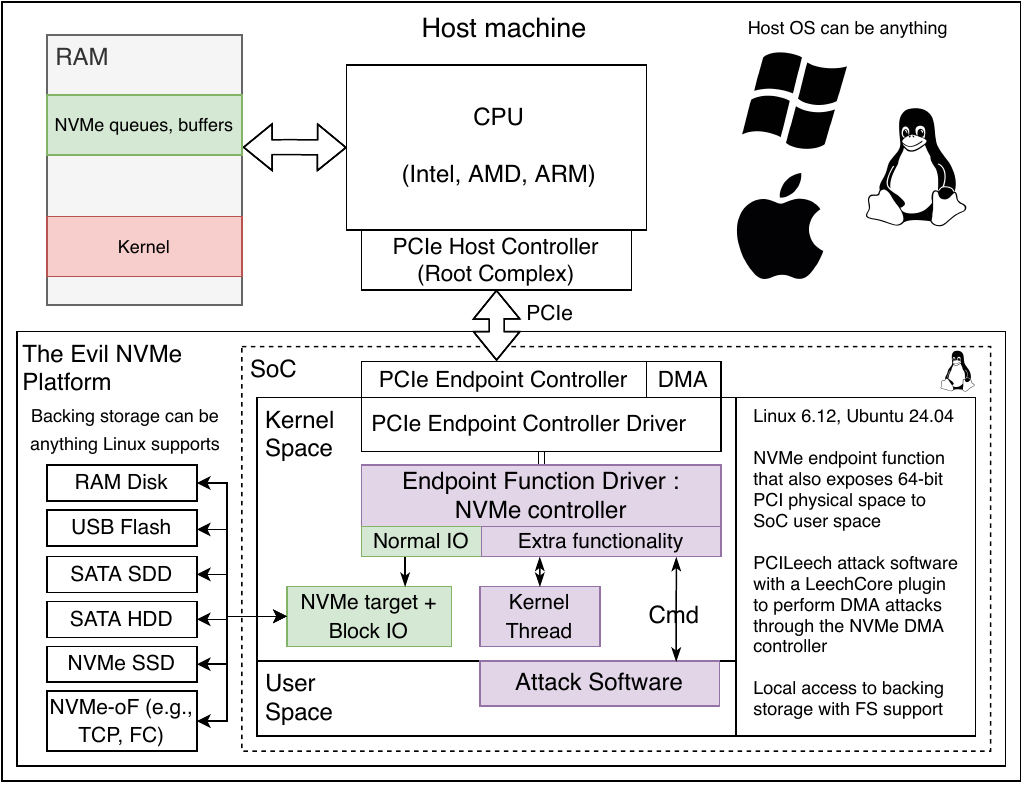}
\end{center}
\caption{\label{fig:csd} \textbf{Architecture of our eNVMe platform}. The NVMe firmware is implemented as an endpoint function that takes care of the normal IO redirected to a storage backend and provides means to execute extra functionality in kernel or user-space.
The fully customizable root file system of the SBC (not visible to host) allows to take advantage of numerous cyber-security software toolchains to probe the target host through PCIe, as well as any file stored on the eNVMe itself.}
\end{figure}

For our implementation, we chose a Rockchip RK3588 SoC based single board computer, the FriendlyElec NanoPC-T6 as shown in Figure~\ref{fig:t6}. The NanoPC-T6 was chosen because of the relatively low cost (sub-200\$), powerful CPU (octo-core ARM), and fast PCIe 3.0 embedded DMAs (up to 4~GB/s).
The RK3588 also provides multiple PCIe controllers\footnote{One PCIe 3.0 x4 dual mode controller (root complex and endpoint), one PCIe 3.0 x2, and three PCIe 3.0 x1 which only function as root complexes.}. As the FriendlyElec NanoPC-T6 was designed to function as a root-complex, for example to plug-in an NVMe SSD. To use it as an endpoint, it is needed to adapt the female M.2 PCIe slot into a male M.2 or regular PCI connector. This can be achieved at very low-cost through off-the-shelf GPU mining PCIe risers with M.2 form factor. These PCIe risers rely on an USB3 cable to carry the PCIe high speed differential signals but allow only for a single PCIe lane. To overcome this limit, we designed our own adapters to change the female M.2 PCIe port into a male M.2 connector. Our adapter split the M.2 PCIe 3.0 x4 onboard female connector into two PCIe 3.0 x2 interfaces. The first one is then connected to a PCIe endpoint controller of the SoC, and the second one is connected to the PCIe root complex controller. The PCIe 3.0 x2 endpoint interface is extended over a mini display port cable to a male M.2 PCB to plug into the target computer. An M.2 female connector on the second interface allows to connect a regular NVMe SSD to the SBC. We chose this in order to benefit from both the endpoint and root-complex capabilities of our SBC and allow to use a regular NVMe as our storage backend. This setup is depicted in Figure~\ref{fig:t6}, besides this setup, it is also possible to use the full PCIe 3.0 x4 in endpoint mode, albeit without the regular NVMe SSD. To connect to the host computer in this mode we rely on a M.2 to regular PCIe adapter and a male-to-male x4 PCIe cable. In this scenario, another backend, e.g., an SD card, eMMC, or a RAM disk has to be used for storage. Our platform allows for any storage backend supported by Linux as shown in Figure~\ref{fig:csd}.

We run a fully open-source software stack composed of a custom 6.12 Linux kernel\footnote{Patches applied to RK3588 drivers, the PCI endpoint framework, NVMe, and the NVMe endpoint function. All available at$^1$.} and an arbitrary root file system (RootFS). For simplicity we chose to install a Ubuntu~24.04 RootFS, but any RootFS can be chosen. The architecture of our Linux-based NVMe SSD is depicted in Figure~\ref{fig:csd}. We are currently working on upstreaming the patches to the PCI endpoint framework and (non evil) NVMe firmware to the mainline Linux kernel. Because the SBC is seen by the host computer as a regular NVMe SSD, and also functions as such, there is no need for any extra driver or changes in the target PC. On the SBC, acting as a man-in-the-middle, we can monitor all of the NVMe requests and traffic. Since the SBC runs Linux, our NVMe drive can also be fully aware of the file system it stores (unless it is encrypted outside of the disk). The SBC can issue PCIe Transaction Layer Packets (TLPs) to read and write the target PC physical address space to carry out DMA attacks.

As said before, the NVMe functionality is implemented as a Linux PCI endpoint function. The Linux PCI endpoint framework allows to write PCI endpoint functions that run within a Linux kernel and are hardware independent thanks to PCI endpoint controller drivers. The endpoint function creates a Linux NVMe target that can make use of any storage backend. In the configuration presented here, the endpoint function code runs as a man-in-the-middle between storage and host computer. Within the endpoint function, in purple in Figure~\ref{fig:csd} we can run extra code either directly or as a kernel thread. This code allows to monitor transfers, alter data, and communicate between user-space and kernel-space. If needed, NVMe commands themselves, with their associated data, can also be passed through user-space via queues implemented as character device interfaces. This flexibility enables running custom code where it makes more sense. Furthermore, not limiting code to kernel space removes the constraint of writing only C code. Access to user-space opens the door to any programming language and software stack the researcher may prefer for the task. User-space also allows launching commands with ease to, for instance, mounting the file-systems of the storage backend on the eNVMe platform. We also expose the entire host physical space to user-space through special a character device, IO operations on this device will be converted to PCIe transfers. This device allows any user-space program to read and write the host PC physical space. Internally, read and write commands are converted to PCIe transactions and rely on the DMA engines to transfer data. Access to the host PCI space from the embedded user-space allows to setup advanced attack schemes as will be presented in the next section.

\section{Attacks}
\label{sec:appattack}

To demonstrate the reality behind the scenarios described in section~\ref{sec:caution} and show the possibilities offered by our \mbox{eNVMe} platform, we implemented several storage-related attacks. First, we start by demonstrating how to activate dormant disks remotely.

\subsection{Remote activation of dormant disks}
\label{sec:remote}

\begin{figure*}[ht]
\centering
\includegraphics[width=\textwidth]{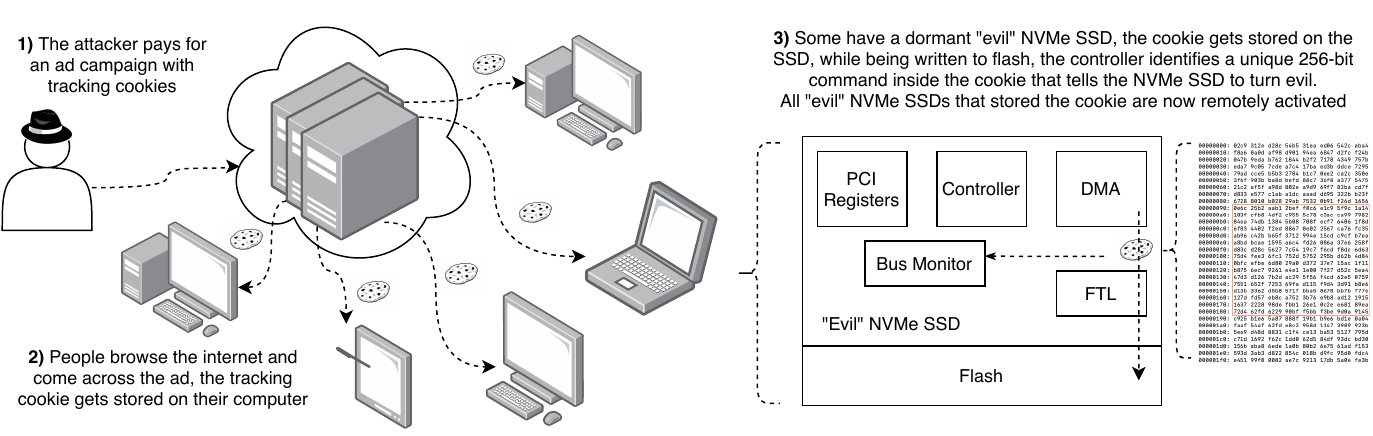}
\caption{\label{fig:activation}\textbf{Remote activation of dormant evil NVMe SSDs}. The attacker uses http cookies that contain hidden commands to remotely activate or control the NVMe SSDs. Tracking cookies through ad campaigns is a widely available tool to get small quantities of arbitrary data written to drives all around the world.}
\end{figure*}

The NVMe drives with dormant malicious firmware cannot be accessed directly by the attacker. However, activation can be done through a side effect. We chose write transactions for this. Our system monitors all disk writes and if the data written contains a pre-assigned pattern (activation key) we enable the dormant features inside our NVMe controller. We chose to have a thread that scans the data transferred in the background as to not impact bandwidth or IOPS of the NVMe as-well-as keep power draw uniform over time. The pre-assigned activation key can be of any chosen length, but should be long enough to avoid spurious unwanted activation through random collisions. If more involved orchestration is needed multiple keys can be used for different actions. As an example of an activation strategy, we chose web HTTP \textit{cookies} as illustrated in Figure~\ref{fig:activation}. \textit{Cookies} are widely used for tracking in web ad campaigns. Because of this they get stored on a large number of computers browsing the web, even if the target computers only browse \textit{legitimate} websites. Therefore, this approach allows to activate a large number of dormant disks with minimal infrastructure and cost simply by running an online ad campaign. As the activation only depends on the activation key being written to the disk, we can also rely on other distribution vectors. For example, regular e-mails, system logs, or cached data from an online source. In summary if the attacker can trigger a write on the remote target system with the activation key inside the data, the dormant disk can be activated. To reduce the risk of collisions even further, a sequence of keys could be required prior to activation. On our SoC we chose to implement the "Bus monitor" of figure~\ref{fig:activation} with a background thread but as we also support FPGA based platforms this could be implemented in hardware at very low power and without any impact on performance. Typically, in a real-world malicious NVMe this could be implemented in-silico as a small state machine that compares the data passing through with minimal impact on power. We implemented the described remote activation and this allows us to trigger any of the attacks described below.

\subsection{Taking control of the host system}

By the fact that we launch the attacks from an NVMe drive we have access to two main areas. First, we have access to all the data stored inside the drive itself. Second, we have access to the PCI space through DMA, which includes the system RAM. Both can allow to take control of the host PC.

\subsubsection{Taking control through the file system}
\label{sec:controlfs}

In the case the host system operating system (OS) is installed on the evil NVMe we can read it and modify it as our NVMe controller has file-system awareness. Most of consumer computers have a single NVMe SSD and the OS is installed on it. With our platform we can modify any part of it, for example: boot partitions, executables, and files. A simple approach would be to replace the bootloader or kernel, however these particularly sensitive parts could be protected for example with secureboot~\cite{wilkins2013uefi, yao2017tour}. Please note that most other executables and files are, in general, not signed or verified.

As a proof of concept we implemented an attack where the eNVMe SSD detects if the OS is Linux-based and if the kernel launches \texttt{/sbin/init}, also called \textit{init} as its first user-space process after boot. \textit{init} coordinates the rest of the boot process and configures the environment for the user once the kernel has booted. \textit{init} runs with \textit{root} privileges in user-space and if compromised allows to take over the system. For our attack, we mount the disk partitions internally and if the OS relies on \texttt{/sbin/init} we replace it with a compromised version. In order not to corrupt the file-system when mounting it locally, it should not be in use by the host computer. We can detect this thanks to the NVMe standard shutdown procedure. NVMe dictates the NVMe device should be disabled by the host and notify the host back that it has shut down correctly before the host completes a total shutdown of the system. This requirement leaves us a perfect opportunity window to access the host file-system without the risk of corruption because we know the host is shutting down and it has already unmounted the file-system. Of course this window is relatively short and should not reach the host timeout values as not to rise suspicion. Once the file replaced, we unmount the file-system and indicate to the host we have shut down correctly. On the next boot the host executes the compromised \textit{init} executable and we can take control. Although this attack works, it modifies the file-system and therefore is quite detectable and leaves the modified \textit{init} executable in-place allowing for analysis and forensics. Therefore, we refined the attack as follows: We mount the file-system as read-only, which we can do anytime during execution. With the file-system mounted we can search for \texttt{/sbin/init} and deduce its physical location on the NVMe. For this, we first locate the position of \textit{init} within the partition with \textit{filefrag}. Then, we compute the partition offset relative to the base logical block address of the NVMe disk with \textit{fdisk}. We now know where this executable is stored and its length can be queried with a simple \textit{ls} command. With this information we proceed as follows: Upon the first NVMe read to the physical location of the \textit{init} executable we transfer our own malicious executable instead of the legitimate file, however, for all subsequent NVMe reads we provide the legitimate data. This allows us to force the host system to execute our own code. Our code will then execute at \textit{root} level before any other process. With this we can inject a kernel module in RAM to get kernel level privileges. Once the payload executed we tell the kernel to drop caches and reread the \textit{init} location, upon which the eNVMe will provide the legitimate data, and we execute the actual \textit{init} executable with \textit{execvp()} so that it is executed with the correct process ID. This makes the detection and analysis of our executable difficult on the host system. Naturally, reading the eNVMe disk from another host or without booting from the compromised partition would reveal our data on the first read of the \textit{init} executable. 

Although we did not go further in obfuscating the injection of a compromised \textit{init} executable we have observed very recognizable read patterns upon system boot e.g., read of the bootloader, then the kernel, the mounting of the file-system and inode reads. As this behavior is consistent and recognizable the eNVMe disk could detect boot read patterns and only deliver the compromised executable in place of \textit{init} in those cases. This would make analysis very difficult as the eNVMe SSD will always provide the legitimate data, e.g., when the disk is read without booting from it, and only provide our compromised \textit{init} in case of a real boot. The eNVMe platform makes it possible to experiment and implement these kinds of advanced scenarios.

With this approach we managed to run a payload with \textit{root} privileges and the following actions are the same as for any rootkit and malicious code that gets executed by a targeted machine. We would like to stress that \textit{init} is only a good example of this strategy, but any other executable or file can be modified on the fly by our attacker eNVMe.

\subsubsection{Taking control through PCI space}

The NVMe SSD can access PCI space which includes the RAM of the target host system. With RAM access we can monitor the target OS, and since we can issue writes to RAM we can inject code in the kernel. We used our PCI DMA capabilities to scan PCI space including the system RAM and locate the host system kernel (both Linux and Windows). The NVMe SSD is allowed to do DMA because this is how NVMe functions. In the case RAM access is limited through IOMMU countermeasures we demonstrate possible ways to circumvent them in section~\ref{sec:noaccess}. Once the kernel is found, we inject a kernel module in RAM and verify its execution by reading RAM. Once we injected the module, we have full control over the target OS. It is interesting to note that the NVMe SSD has even higher access privileges than the \textit{root} user has: the NVMe can access all physical addresses of the RAM whereas, for instance, under Linux, the \textit{root} user cannot access physical RAM by default, unless the Linux kernel is explicitly configured to allow it. We implemented the DMA attacks and kernel module injection by porting the open-source PCILeech\footnote{\url{https://github.com/ufrisk/pcileech}} Direct Memory Access (DMA) attack software~\cite{frisk2016direct, frisk2016dma, frisk2018rise} to our eNVMe platform and interfacing it with our NVMe firmware. Once the kernel module loaded in the host RAM we have a fully functional \textit{root} console that is accessible from within our eNVMe SSD to issue commands. On both attacks either through the file-system or through PCI space we gain total control over the host system from within the NVMe SSD.

\subsection{Root Shell}

Gaining access to a root shell is a priority for any attacker. Through the injected kernel module, we can spawn a root shell (or system shell in Windows) and execute any commands with root privileges. This allows for an easy pathway to implement further attacks. Through shell commands the eNVMe SSD can change passwords, move files, ping servers, download malware, spawn reverse SSH connections, and do whatever an attacker can do with a root shell. The possibilities open to the eNVMe are endless. This is also the easiest way for the eNVMe to move data back to the attacker, for example stolen crypto-wallets or authentication keys.
Against an Ubuntu 22.04.3 LTS target with a Linux 6.4.0 kernel and 64~GB of RAM on a Ryzen 3900X CPU and X570 chipset, our eNVMe SSD scanned the memory and injected a kernel module in 42 seconds. Once the module loaded, we can spawn a root shell and execute arbitrary commands. Against a Windows 10 Enterprise running on the same machine, our NVMe SSD achieved the same in 11 seconds.

\subsection{Accessing the target from the embedded Linux}

Through the injected kernel module, we are able to map the entirety of the target system RAM and file systems into the Linux system embedded in our eNVMe platform. We can now, from within eNVMe, open the host RAM as a memory mapped file and execute read-write instructions as we please. This allows for running numerous memory analysis tools directly from the eNVMe user-space. Note that, for this, the kernel injection is not even a necessity if we can access the RAM through DMA directly. However, in cases where the IOMMU blocks us, but a kernel module was injected through the file-system, this mapping is of interest. The kernel module also helps us by using the target OS API to query the RAM size, mappings, and boundaries, which we cannot get directly from PCIe reads and DMA access. With the injected kernel module, we can also mount other target's file-systems (resident on other storage devices or remotely) inside our embedded Linux. As an example, on a target machine running Windows, installed on a different regular NVMe SSD that is not under our control, we can still access all the files through this mount point, through the injected kernel.

\subsection{An army of disks}

Remote activation, e.g., through cookies, also allows to control the disks, giving them instructions. As this may not be the most effective way to communicate, all activated disks that have control over their host machine could open other channels to the attacker to get instructions instead. This could be remote reverse shells or download of further instructions from any server. Because the NVMe SSD has control of the host machine memory space, the NVMe SSD can become the de-facto brain of the infected machine and be used to create a new type of botnet. Such weaponized NVMe SSDs could have colossal consequences if deployed widely.

\subsection{Denial of Service}

Denial of Service (DoS) attacks can also be implemented through remote activation. On reception of the activation key the NVMe SSD can operate different DoS attacks. If the NVMe has control over the host anything is possible, but even without control of the host PC, the NVMe itself can chose to stop functioning, provide corrupted data, or to an extreme self-destroy. If the NVMe drive is the host PC main drive the PC is now non-operational. Thanks to the distributed nature of the remote activation a large scale Distributed Denial of Service (DDoS) attack can be accomplished with large tactical or economical impacts similar to the \textit{Crowdstrike} outage~\cite{george2024trust, demicrosoft}. There is another low hanging fruit accessible to eNVMe besides self-destruction: auto encryption. If the disk decides to cipher its content, it can easily act as a ransomware would.

\subsection{On-line data mining and extraction}
In the \textit{apocalypse} section we describe that the NVMe SSDs could be used for spying, extraction of files such as \textit{ssh-keys} or \textit{cryptowallets} or any other kind of intelligence. We have shown that the eNVMe SSD can self-mount the partitions written to it in section~\ref{sec:controlfs}, this can be used to scan for certain documents and easily allow extraction of files stored at typical locations e.g., \texttt{.ssh}. We chose to push the experiment further and show that we can use embedded artificial intelligence (AI) technology to collect data based on specific features.

\subsubsection{AI-based data mining}

As a proof of concept of AI-based data mining inside the eNVMe SSD we chose to mount the host file-system internally as read-only, again to avoid corruption, and used AI analysis of typical folders that would hold pictures (e.g., \texttt{/home/user/Pictures}) to detect and extract pictures of interest.
We tested this approach with the OpenAI CLIP (Contrastive Language-Image Pre-Training) neural network~\cite{radford2021learning}. This network is trained on text-image pairs and can compute distances between images and text snippets in an embedding space. This allows to search for image with text or search images for similarities with a given image. We setup the \mbox{ViT-B/32} pre-trained model on our eNVMe platform, details and hyperparameters are shown in Table~\ref{clip_hyp}. This model is relatively lightweight (338~MB) and allows the eNVMe SSD to use CLIP vision transformer inference to compute embeddings of any picture stored. Based on distance in the embedding space any pictures with the desired features can be extracted. We chose to implement this approach to show that it is possible to do low-power AI inference on an embedded system with malicious intent.

\begin{table*}[ht!]
\caption{CLIP (Contrastive Language-Image Pre-Training) Model Vit-B/32 hyperparameters. Context length:~77, Vocab size:~49,408. Pretrained model from~\cite{radford2021learning}, available through \url{https://github.com/openai/CLIP}.}
\label{clip_hyp}
\centering
\begin{tabular}{l|ccccccccc}
\hline
\multirow{2}{*}{Model} & \multirow{2}{*}{\begin{tabular}[c]{@{}c@{}}Size\\ {[}MB{]}\end{tabular}} & \multirow{2}{*}{\begin{tabular}[c]{@{}c@{}}Embedding\\ Dimension\end{tabular}} & \multirow{2}{*}{\begin{tabular}[c]{@{}c@{}}Input\\ Resolution\end{tabular}} & \multicolumn{3}{c}{Vision Transformer} & \multicolumn{3}{c}{Text Transformer} \\
                       &                                                                          &                                                                                &                                                                             & layers      & widths      & heads      & layers      & widths     & heads     \\ \hline
ViT-B/32               & 338                                                                      & 512                                                                            & 224                                                                         & 12          & 768         & 12         & 12          & 512        & 8        
\end{tabular}
\end{table*}

\begin{table*}[ht!]
\caption{CLIP vision transformer compute time on the RK3588-based eNVMe platform per batch of $N$ new images to generate the embeddings. Batches of new images are randomly sampled from the COCO dataset~\cite{lin2014microsoft} (123,403 .jpg images, $\approx 20$ GB).}
\label{clip_inf2}
\centering
\begin{tabular}{ccc|ccc|ccc}
\hline
\multicolumn{3}{c|}{$N=100$ new images}                                                                                                                                           & \multicolumn{3}{c|}{$N=1,000$ new images}                                                                                                                                         & \multicolumn{3}{c}{$N=4,000$ new images}                                                                                                                                          \\
\begin{tabular}[c]{@{}c@{}}Size\\ {[}MB{]}\end{tabular} & \begin{tabular}[c]{@{}c@{}}Embedding\\ Computation\end{tabular} & \begin{tabular}[c]{@{}c@{}}time/\\ image\end{tabular} & \begin{tabular}[c]{@{}c@{}}Size\\ {[}MB{]}\end{tabular} & \begin{tabular}[c]{@{}c@{}}Embedding\\ Computation\end{tabular} & \begin{tabular}[c]{@{}c@{}}time/\\ image\end{tabular} & \begin{tabular}[c]{@{}c@{}}Size\\ {[}MB{]}\end{tabular} & \begin{tabular}[c]{@{}c@{}}Embedding\\ Computation\end{tabular} & \begin{tabular}[c]{@{}c@{}}time/\\ image\end{tabular} \\ \hline
16                                                      & 1 min 12 s                                                      & 0.72 s                                                & 158                                                     & 9 min 39 s                                                     & 0.58 s                                                & 628                                                     & 38 min 22 s                                                     & 0.58 s                                               
\end{tabular}
\end{table*}

We measured inference time on batches of new images to assess the feasibility of on-line AI-based data mining in our eNVMe platform. The results are shown in Table~\ref{clip_inf2}. The CLIP model runs on the RK3588 CPUs without any acceleration and can compute embeddings for a batch of 1,000 new images in less than 10 minutes. We also measured the power consumption difference between inference and idle operation. When idle the eNVMe platform draws just over 4~watts of power, when we run inference with CLIP the power draw increases by 2~watts, a 50\% increase. Typical power draw of an NVMe SSD is between 1-3~W in idle and can reach above 10~W under load, with a maximum of 25~W\cite{rana2024ssd, nvme2024power}. The model and runtime parameters can be further tuned to reduce the power draw, even at the expense of runtime. Furthermore, it would be possible to improves this through hardware acceleration, as an NVMe designed for this might include a small neural engine in its chip. These scenarios can also be studied with the chosen hardware platform considering that the RK3588 provides a Neural Processing Unit (NPU) as well as a Mali GPU.

A similar approach could be taken to identify and extract target text documents with Large Language Models (LLMs). LLMs could be used to detect documents of interest but would also allow to summarize contents and generate lower size summaries of user stored data for extraction. Performance-wise, inference time is not necessarily critical as the NVMe SSD is usually on for extended periods of time without major new file writes and not all files need to be scanned at once. Therefore, AI based scans can be tailored to be done at lower-power to avoid giving rise to suspicion. Tagged files can then be copied from the file-system to a private internal memory of the NVMe drive for collection to be extracted at a later time either through control of the host system, network card, or other side-channel communication channels.

\subsection{Side-channel communications from NVMe}
\label{sec:side}

Other communication channels from the NVMe SSD to the outside world are of interest if the NVMe SSD cannot take control over the host target system. There are multiple possibilities. First, it can read and write other PCIe devices (if the IOMMU is enabled at least the ones inside the same IOMMU group). Through PCIe the NVMe SSD can hijack networking devices and establish communication to the outside world without even going through the host OS~\cite{kristiansen2015pcie}. Other means of communication can be through radio communication~\cite{barbeau2006detection, liu2018covert} and sonic side-channels~\cite{arp2017privacy, yan2020surfingattack, duan2021privacy}. As a demonstration we implemented a simple Wi-Fi side-channel from within our eNVMe platform. On real malicious hardware, the NVMe SSD could have an antenna to communicate through radio frequencies directly embedded in the PCB or a piezo element on the PCB and communicate through sonic or ultra-sonic waves to other compromised nearby devices. One particularly interesting side-channel here is LoRa (Long Range) communication~\cite{augustin2016study, bor2016lora, devalal2018lora} with an antenna embedded in the NVMe PCB. This allows for silent transmission, i.e., low-power, of small packets over large distances -- perfect for extraction of passphrases or private keys stored on the drive.

\subsection{NVMe spoofing}

The PCILeech DMA attack software~\cite{frisk2016direct, frisk2016dma, frisk2018rise} states one of the main limitations as "Recent Windows and Linux versions block DMA by default"\footnote{\url{https://github.com/ufrisk/pcileech?tab=readme-ov-file\#limitationsknown-issues}}. DMA here references PCI devices with the DMA class code\footnote{Base class 0x08, Sub-class 0x01 (DMA controller).}. Thanks to our NVMe security research platform we discovered that if a PCIe device uses the class code for NVMe\footnote{Base class 0x01, Sub-class 0x08, Programming interface 0x02, (NVM Express (NVMe) I/O controller).} and implements PCIe BAR0, Linux (as of 6.8) and Windows 10 will still recognize it as an NVMe SSD even with BAR0 registers only reading 0's and will still allow the device to remain enabled. Both OSes report an issue with the disk, Linux as the disk not becoming \textit{ready}, and Windows as the disk not being able to be configured correctly. However, neither disables the PCI device nor unloads the associated driver. Thus any PCI device spoofing its class code to an NVMe I/O controller is not blocked as a device with DMA class code would be by the target OS and thus remains enabled\footnote{We are working on patches to the Linux NVMe driver to change this behavior.}. This example shows the value of our eNVMe platform as it allows to probe many parts of the host system that for the moment are mainly unexplored. Even with such simple questions as \textit{What would happen if we tell the host system we are an NVMe device but never become ready?} Such questions arise all over at any level in the NVMe protocol and specifications.

\subsection{Further attacks}

We have demonstrated the reality of several attacks that could be performed by a malicious NVMe device and our eNVMe platform allows researchers to extend the range of attacks with relative ease thanks to our software-based approach with support for a full-fledged Linux OS inside our eNVMe device. The low-cost and availability of the required hardware also makes it reachable and invaluable tool for any security researcher with interest in NVMe.

\section{Mitigations}

Several mitigations and countermeasures can be deployed to limit the reach of the different attacks presented so far. This section presents them and discuss their benefits and limitation as well as how to bypass them when possible.

\subsection{IOMMU}
\label{sec:iommu}

IOMMU is the most important mitigation against DMA attacks. It allows to isolate the NVMe SSD and only let it access part of the physical memory space under the control of the host OS, similarly as to how the MMU isolates processes in a virtual address space. This is illustrated in Figure~\ref{fig:nvme1} where the NVMe is only allowed access to the green regions.  Without IOMMU the NVMe SSD, or any other DMA-capable device, can simply take total control of the host computer because it can read and write any location in the physical memory. Most modern CPUs and chip-sets are equipped with an IOMMU, but unfortunately, it is not enabled or properly configured on a large number of  machines~\cite{markettos2019thunderclap}. Furthermore, even after years from this survey, the situation has not improved significantly. Activating IOMMU by default generates many user-visible miss-behaviors, most probably due to driver code, that are complex to solve. To this date many Windows versions and Linux distributions require manual user activation and sometimes complex procedures to enable and configure the IOMMU. Emblematic on this respect, is the discussion around the default settings for the Ubuntu distribution that, after having opted for a default active IOMMU, revised this decision due to problems with the graphics stack~\cite{ubuntu2022dma}. The Fedora distribution has the IOMMU disabled by default and requires manual intervention to enabled it~\cite{fedora2024hard}. BIOS software is also sometimes shipped with IOMMU disabled by default and requires user intervention to enable it~\cite{iommu2015page}. This means that, even if the OS comes with IOMMU drivers and is correctly configured, BIOS settings prevails and the IOMMU will not be used. 

IOMMUs, as other MMUs, works with a page granularity. The implication of this basic fact is subtle: if no special care is taken in kernel and driver code, these page sized mappings could leak information. This mix of sharing data and other information in the same DMA area has been studied for Network Interface Cards (NICs) and lead to many successful attacks\cite{markettos2019thunderclap}. Our platform enables this kind of investigations for the storage sub-system in modern operating systems.
Another intrinsic limit of IOMMUs is that they cannot isolate PCI devices which are part of the same IOMMU group, e.g., under the same PCIe switch. Peer-to-peer attacks between devices on the same bus are not new~\cite{sang2011attacks}, and research is ongoing to provide more isolation between devices on hardware and virtualized PCIe busses~\cite{jang2019heterogeneous, smolyar2015securing}. If the NVMe SSD and a Network Interface Card (NIC) are inside the same IOMMU group, an IOMMU cannot stop the NVMe SSD to interact with the NIC~\cite{redhat2024deep}. Motherboards manufacturers are usually terse in the amount of information they share in their documentation about this specific point. Due to the lack of PCIe lanes on a CPU, chipsets on the motherboard often provide extra PCIe lanes through a bridge and these extra lanes (for M.2, audio devices, network bridge, and cards) are grouped together. Therefore, security-concerned users have to go through a painful trial and error process to assess the grouping granularity of a system. Finally, for performance reasons, IOMMU mappings are cached as it is the case for MMUs in TLBs. These caches are not immune to temporal side channels as demonstrated by~\cite{gras2017aslr, koschel2020tagbleed, kim2023devious}.

\subsubsection{Defeating the IOMMU}
\label{sec:defiommu}
\label{sec:noaccess}
Notwithstanding the limitations mentioned above, the IOMMU remains the best protection against DMA attacks. Acting as an NVMe drive gave us the unique opportunity of devising some specific strategies that can be deployed to disable a properly configured IOMMU.
As a proof of concept, we setup the eNVMe drive so that it detects if a Linux OS with an active IOMMU was installed in the drive. As said, a Linux installation making use of the IOMMU would not let our NVMe drive take full control of the machine through PCI space. However, our NVMe SSD is aware of what is written on it, so it patches the Linux installation and adds the kernel flags to disable IOMMU. Upon next reboot of the machine, the Linux kernel loads without activating the IOMMU, and our NVMe drive now can take full control of the kernel by accessing the RAM. This attack relies on the fact that the RootFS stored on the NVMe is not encrypted outside of the NVMe drive and therefore the controller can read it. To achieve this, we scan the file system to check if it contains the popular GRUB bootloader. When found, we update the \texttt{/boot/grub/grub.cfg} file so that the kernel \texttt{boot args} \texttt{GRUB\_CMDLINE\_LINUX\_DEFAULT=} contain \texttt{amd\_iommu=off}, and \texttt{intel\_iommu=off}. The next time the kernel boots, the IOMMU is disabled and we can take control of the machine. Because this change is visible in the file and can be obvious, we can apply the same strategy described in section~\ref{sec:controlfs} and only provide the modified file when we detect the machine is actually reading it for booting. Another similar possibility is to directly patch the \texttt{vmlinuz-} kernel image in the boot directory. There exists also another opportunity to take control of the host RAM contents, hibernation. Hibernation, also known as suspend to disk, is powering down a computer while retaining its state. When hibernation begins, the computer saves the contents of RAM to the storage medium. Windows compresses and writes the RAM contents to the \texttt{hiberfile.sys} file and of which contents can be extracted~\cite{mrdovic2011forensic, kost2019extract, mosse2022windows}. On Linux hibernation relies on a swap file or partition, and similarly it can be extracted~\cite{nikkel2021practical, swsusp2bin}. Therefore, if the storage medium is the eNVMe, and the partition holding the file is not encrypted, the eNVMe could scan the contents of the RAM being stored and patch them on the fly to inject a kernel module, when the systems resumes and restores the RAM contents from the eNVMe, it loads the malicious kernel module itself. Hibernation file forensics research has shown that this approach allows to extract and modify critical information~\cite{murtuza2015tool, ayers2015windows, ghafarian2020windows}, with the benefit that hibernation file attacks are not affected by the IOMMU at all.

Furthermore, for cases when deactivating the IOMMU is not directly possible, there still exist multiple, more involved, attack paths to circumvent it~\cite{wojtczuk2009another, sang2010exploiting, wojtczuk2011following, jacob2017break, kupfer2018iommu, morgan2018iommu, gross2022breaking, kim2023devious}.

\subsection{Disk encryption}
In our attacks, the disk itself can read its content and extract valuable information as well as modify data stored in it or provide modified data on specific reads. This is a very potent and orthogonal attack vector that complements DMA exceptionally well as shown in section~\ref{sec:defiommu}.
If data are encrypted, many attacks we present previously become almost impossible. However, hardware full-disk encryption (FDE)~\cite{fruhwirth2005new, codandaramane2016securing, khati2017full}, a complex technology, plagued by many shortcomings~\cite{lopez2013studies, muller2014systematic}, cannot protect the data in the considered scenario. This is because our disk (ill-intended) firmware holds the encryption keys in a FDE scenario nullifying encryption benefits. Even when acting as man-in-the-middle with a fully encrypted backend SSD (our device doesn't hold any encryption key), our controller still sees unencrypted data: FDE only protects data at rest and not in transit. All the exchanges of data between host and controller are perfectly visible to our \textit{evil} NVMe device. 

To be effective, encryption has to happen outside of the \textit{evil} NVMe disk and before transmission to the device. This can be accomplished through software solutions such as \textit{BitLocker} on Windows~\cite{microsoft2024bitlocker} and \textit{dm-crypt}, the disk encryption subsystem in Linux~\cite{dmcrypt}. However, this solution adds another step of processing before data reaches the disk impacting system load and performance. Reports indicate that software encryption burden results in only 50\%-80\% of the original performance~\cite{walton2023tested, larabel2018cost, korchagin2020speeding, del2023perfluks} and is therefore not used in general. Furthermore, software encryption is difficult for boot devices because not all BIOS and bootloaders fully support a software encrypted disk. On top of that, in our threat model, the encryption engine and keys must be loaded prior to boot from a different and trusted storage to decipher the disk contents. If the keys are stored on the NVMe device they are accessible. In the case they are passphrase protected, the NVMe device could still try finding the passphrase over long periods of time. Nevertheless, a promising approach solving many of the mentioned issues has been recently presented \cite{Chowdhuryy2023d-shield} and we hope some similar schemes will be largely adopted soon.

\subsection{Minimal set of required mitigations}

Full protection against a rogue disk is impossible, but we can mitigate the risks as much as possible. Table~\ref{table:min} lists the minimal set of required mitigations to protect against the attacks we described in this work. 

\begin{table}[htbp]
\caption{Minimal set of mitigations to shield from evil NVMe drives.}
\label{table:min}
\begin{center}
\begin{tabular}{|l|p{45mm}|}
\hline
\textbf{Mitigation}   & \textbf{Reason} \\
\hline
IOMMU                 & IOMMU is a necessity to isolate any DMA-capable device. Without it, the device can take total control.                  \\
\hline
Secureboot            & Secureboot allows to check that both the boot loader and kernel are properly signed and have not been tampered with.    \\
\hline
Off-disk cryptography & If the disk can read its own contents for ill-intent, data accessible to the disk should be indecipherable by the disk. \\
\hline
\end{tabular}
\label{tab1}
\end{center}
\end{table}

Off-disk cryptography should almost totally protect against remote activation; however, due to collisions, the activation key pattern could still be written to the drive by chance. Nonetheless, due to the long nature of the binary key (e.g., 1024-bits or more) the probability of this happening is quite low, and with the other protections in place the most the drive can do is self-destructive actions (destroy or alter stored contents or itself). This would be similar to traditional hardware faults and failures occurring in storage drives. For which the mitigations would be a regular backup scheme. Even when the presented activation method is impossible, there are alternative activation techniques. For instance, a large number of disks could be programmed to disable them selves simultaneously at a given time, chosen from factory or by firmware update, creating a much harder situations to handle, with consequences similar to the recent \emph{Crowdstrike}~\cite{george2024trust, demicrosoft} failed update which created a large scale disruption. Remote activation could also be triggered by other side-channels as discussed in section~\ref{sec:side}. Protecting against activation from hidden side-channels is a difficult and costly task, but not an impossible one.
Secureboot allows to sign and verify important parts of the boot process such as the bootloader and kernel image, but it also comes with limitations, for example it makes it impossible to load modules that are not signed by a trusted key, by default this will block all out-of-tree modules, including DKMS-managed drivers (e.g., closed source GPU drivers)~\cite{debian2024wiki}. On Linux systems it can even disable hibernation totally~\cite{debian2024wiki}. When more than one disk is present in the system, a redundant array of independent disks (RAID), configured to distribute data, could be a reasonable way to avoid paying the encryption penalty while at the same time not showing all the file-system data to an \textit{evil} device. This mitigation is, however, not applicable to all systems, laptops with a single NVMe device come to mind. Nonetheless, in multi-disk settings, a RAID with multi-vendor disks could represent a mitigation denying access to file-system and files content from \textit{eNVMe}.

The main problem remains, and does not come from the lack of mitigation techniques, the problem is the fact that: most of them are not enabled by default, or impractical to be fully implemented at large scale. If a consequent number of computers hosting a malicious NVMe device do not implement these mitigations, which is the case in the current computing landscape, weaponized NVMe SSDs remain a major and under-studied threat. Also, it is not realistic to think that most user grade computers will implement these mitigations within the next several years, making this threat very concrete. The lack of research around harmful NVMe devices makes it difficult to assess if such devices exist already however within this work, we show that the creation of weaponized NVMe SSDs is possible and we hope to provide the groundwork to study further NVMe based attack scenarios.

\section{Discussion}

Because of the immense potential impact of weaponized storage, difficulty to setup effective mitigation techniques that cover all cases at scale, and complexity in examining existing disks for risk of threats, we have to think of alternative solutions to the problem. An optimistic solution would be \textbf{transparency}. With transparency, as in open-source software and open-hardware, security experts can inspect and check storage controllers and firmware to make sure that they are not \textit{evil}. Although even with open solutions malicious code can still be inserted~\cite{chin2021how, Lakshmanan2024malicious}, this malicious code often comes from sources external to the project. Publishing open-source malicious code, even if obfuscated, would come with great risk to reputation and trust. Relying on trusted closed-source NVMe drives does not protect against malicious functionalities that could be activated in extreme-cases. Even worst, nothing prevents the general public from buying NVMe drives from untrusted sources, for example if it makes economic sense. Even when we trust our supplier, regular NVMe SSDs could get reprogrammed with malicious intent somewhere in the supply chain or via compromised firmware update at a later time. Attempts to prevent these issues exists, such as Microsoft Device Identifier Composition Engine (DICE)~\cite{microsoft2024dice}. DICE allows for hardware-based cryptographic device identity, attestation, and data encryption. 
A DICE identified device makes it harder to tamper with signed firmware or device identify because the attacker does not possess the private keys to do so. While such technologies exist, few systems actually use it. In all cases black box implementations rely on trust, even when signed. While there is no silver bullet, we believe that an open initiative for storage controllers would benefit security and help storage research in general. With the NVMe standards already being open, as well as, the flash side interfaces with the Open NAND Flash Interface (ONFI)~\footnote{\url{https://onfi.org}} also being open, it makes sense to promote an open controller. Flash manufacturers would still be able to produce closed-hardware flash chips, but it is much harder to implement attacks at the NAND level. Open controllers could power novel open-SSDs that would provide an alternative where it makes sense, for example in highly sensitive fields such as governmental, military, or banking sectors. Review and access to the firmware in the end product is a good way to ensure a device is safe. With closed hardware this will remain out of reach, because even with a third-party-reviewed and signed firmware, an \textit{evil} NVMe drive could still present a genuine firmware to the host computer but internally run malicious code. Therefore, there is a strong reason to advocate for transparency in storage devices, and push for open-hardware controllers with open-source firmware.

We think that for institutions threatened by the scenarios demonstrated in this work, adopting open-devices would be a sensible path to follow even if it comes at with an economical and performance impact, albeit a reasonable one. For the moment there are no commercially available open NVMe SSDs, but a few research initiatives exist~\cite{bjorling2017lightnvm, openchannelssd, kwak2020cosmos+, openssd}. None of them produce an end-user product. So, there is a contingency to go towards an open-controller initiative or even an entire open NVMe SSD.

Besides high-impact fields, the large quantity of NVMe SSDs used by ordinary consumers makes a distributed NVMe SSD attacks a very powerful cyber-weapon. 
Globally moving to open solutions is a utopian goal and even if initiatives push towards it, it may never be fully adopted. Therefore, the best approach for the moment is for operating systems and BIOS firmware to enable security features by default, push new installations to enable secureboot, IOMMU, and off-disk encryption by default, providing an opt-out for the user, rather than the opposite currently implemented. It is understandable that these can come with limitations, worst performance, and frustration for the user so it is important to be able to choose a course of action based on actual threats. This requires new research. First, on what is possible with malicious NVMe device. Second, research on the possibility and chances of malicious NVMe devices already being released or created. These research areas are complementary and the first is useful to drive the second. With this work we provide a platform that allows for research and prototyping any kind of attack an NVMe device could be capable of. Thanks to this work researchers now have a new valuable tool to explore and continue paving the way towards more secure storage and safer storage infrastructures.

\section*{Acknowledgments}

This work has been supported by \if\anon1{... (hidden for anonymous review).}\else{the School of Engineering and Management Vaud (HEIG-VD).}\fi

\end{document}